# American Options Pricing under Stochastic Volatility: Approximation of the Early Exercise Surface and Monte Carlo Simulations

Kuperin Yu.A., Poloskov P.A.

Department of Physics, Saint Petersburg State University, Saint Petersburg, Russia

**Abstract**

The aim of this study was to develop methods for evaluating the American-style option prices when the volatility of the underlying asset is described by a stochastic process. As part of this problem were developed techniques for modeling the early exercise surface of the American option. These methods of present work are compared to the complexity of modeling and computation speed. The paper presents the semi-analytic expression for the price of American options with stochastic volatility. The results of numerical computations and their calibration are also presented. The obtained results were compared with results excluding the effect of volatility smile.

## Introduction

The main difference between the American option and the European option is that the American option can be exercised at any time before the expiration. This considerably increases the difficulty of the problem, because now it is not known boundary condition for the time variable. In this situation, one needs to calculate the surface of the early exercise surface. To do this we shall use the method proposed in [11]. Then one can use an analytical expression for the price of the American type option, based on the results obtained in [5].

## The basic model

As in the case of European options, we shall use the geometric Brownian motion as a model of underlying asset dynamics. And to describe the dynamics of the underlying asset volatility we shall choose Heston square root process:

$$dS = \mu S dt + \sigma S dz_1$$

$$dV = k(Q-V)dt + \xi V^{1/2}dz_2$$

Stochastic processes $z_1$ and $z_2$ are correlated with correlation coefficient $\rho$, namely $E[dz_1 dz_2] = \rho dt$. If $r$ is the risk-free interest rate and $q$ is the continuously compounded dividend rate, then applying the Ito Lemma and acting standard methods one can obtain a partial differential equation for the price of American Call option:

$$\frac{\partial C_A}{\partial \tau} = \frac{VS^2}{2}\frac{\partial^2 C_A}{\partial S^2} + \rho\sigma VS\frac{\partial^2 C_A}{\partial S \partial V} + \frac{\sigma^2 V}{2}\frac{\partial^2 C_A}{\partial V^2} + (r-q)S\frac{\partial C_A}{\partial S} + (k(Q-V) - V\lambda)\frac{\partial C_A}{\partial V} - rC_A$$

(1.1)

$$0 \leq \tau \leq T$$
$$0 < S < b(V,\tau)$$
$$0 \leq V < \infty$$

Here $b(V,\tau)$ denotes the early exercise boundary at time $\tau$ and volatility level $V$. The initial and boundary conditions for (1.1) in the case of an American call option are

$$C_A(S,V,0) = \max(S-K,0)$$
$$C_A(b(V,\tau),V,\tau) = b(V,\tau) - K$$
$$\lim_{S \to b(V,\tau)} \frac{\partial C_A}{\partial S} = 1 \qquad (1.2)$$
$$\lim_{S \to b(V,\tau)} \frac{\partial C_A}{\partial V} = 0$$

The right hand side in the first equation (1.2) is a payoff function and may take another form, but in our case, we confine ourselves to plain vanilla payoff function. The first step to solve the problem (1.1) with free boundary conditions (1.2) will be the change of variables. $S = e^x, U_A(x,V,\tau) = e^{r\tau}C_A(S,V,\tau)$. After change of variables equation (1.1) is rewritten as:

$$\frac{\partial U_A}{\partial \tau} = \frac{V}{2}\frac{\partial^2 U_A}{\partial x^2} + \rho\sigma V\frac{\partial^2 U_A}{\partial x \partial V} + \frac{\sigma^2 V}{2}\frac{\partial^2 U_A}{\partial V^2} + (r-q-\frac{V}{2})S\frac{\partial U_A}{\partial x} + (\alpha - \beta V)\frac{\partial U_A}{\partial V} \qquad (1.3)$$

In this equation we introduced, for the convinience, the following variables: $\alpha = kQ$, $\beta = k + \lambda$. The boundary conditions after the change of variables take the form:

$$U_A(x,V,0) = \max(e^x - K, 0)$$
$$U_A(\ln(b(V,\tau)), V, \tau) = (b(V,\tau) - K)e^{r\tau}$$
$$\lim_{x \to \ln b(V,\tau)} \frac{\partial U_A}{\partial x} = b(V,\tau)e^{r\tau} \qquad (1.4)$$
$$\lim_{x \to \ln b(V,\tau)} \frac{\partial C_A}{\partial V} = 0$$

In the article Chiarella, Zuiogaz [9] was proposed a method which allows obtaining the inhomogeneous differential equation for an unbounded domain $-\infty < x < \infty$, which would be equivalent to equation (1.3) for a bounded domain $-\infty < x < b(V,\tau)$ According to [9] the problem (1.3) - (1.4) can be written as

$$\frac{\partial U_A}{\partial \tau} = \frac{V}{2}\frac{\partial^2 U_A}{\partial x^2} + \rho\sigma V \frac{\partial^2 U_A}{\partial x \partial V} + \frac{\sigma^2 V}{2}\frac{\partial^2 U_A}{\partial V^2} + (r - q - \frac{V}{2})\frac{\partial U_A}{\partial x} + (\alpha - \beta V)\frac{\partial U_A}{\partial V} +$$
$$+ H(x - \ln b(V,\tau))\{e^{r\tau}(qe^x - rK)\} \qquad (1.5)$$

where $H(x)$ is the Heaviside function.

This equation should be solved in the unbounded domain $-\infty < x < \infty$, $0 < V < \infty$, $0 \leq \tau \leq T$.
Note that the boundary and initial conditions (1.4) also hold for this equation.

**Approximate solution by means of Fourier transform**

Inhomogeneous equation (1.5) differs from the homogeneous equation, which solved Heston (1993) for European options with stochastic volatility only by term, giving the inhomogeneity. Then, solution of (1.5) can be written as the sum of two terms:

$$U_A(x,V,\tau) = U(x,V,\tau) + W(x,V,\tau) \qquad (1.6)$$

The first term in (1.6) is a solution of the inhomogeneous equation for the European option, and the second one is called early exercise premium [9]. In this case, the second term we will find in the form

$$W(x,V,\tau) = \int_0^\tau \int_0^\infty \int_{-\infty}^\infty H(\zeta - \ln b(w,\xi))\{e^{r\xi}(qe^u - rK)\} F(x,V,\tau-\xi,u,w) du\,dw\,d\xi$$
$$(1.7)$$

The value $W(x,V,\tau)$ represents the expectation of the discounted to zero - time cash flow $(qe^u - rK)$. The expectation is calculated at the asset price $x$, at the volatility $V$ and at the time $\tau$ remaining until the option is exercised. Function $F(x,V,\tau-\xi,u,w)$ is a transitional probability. The function $U(x,V,\tau)$ has already been found in the work of Heston (1993). It means that we have to calculate $W(x,V,\tau)$ only, which in turn requires the computation of the function $F(x,V,\tau-\xi,u,w)$.

Since it is impossible to obtain an analytical solution for $F(x,V,\tau-\xi,u,w)$ at an arbitrary function $b(V,\tau)$, we use linear approximation to the early exercise surface, which was proposed in Tzavalis & Wang (2003). These authors showed that the logarithm of the function $b(V,\tau)$ can be accurately approximated by a linear function of $V$:

$$\ln b(V,\tau) \approx b_0(\tau) + Vb_1(\tau). \quad (1.8)$$

Following [9] we use Fourier transform to obtain an semianalytical expression for $W(x,V,\tau)$.

First, apply the Fourier transform of the function $F(x,V,\tau-\xi,u,w)$ with respect to the variable $u$:

$$\hat{f}_2(x,V,\tau-\xi,\phi,w) \equiv \int_{-\infty}^{\infty} e^{i\phi u} F(x,v,\tau-\xi,u,w)du.$$

Then equation (1.7) can be rewritten as

$$W(x,V,\tau) = \int_0^\tau \int_0^\infty \int_{-\infty}^\infty H(u-\ln b(w,\xi))e^{r\xi}(qe^u - rK) \times \left\{\frac{1}{2\pi}\int_{-\infty}^\infty e^{-iu\phi}\hat{f}_2(x,v,\tau-\xi,\phi,w)d\phi\right\} du\, dw\, d\xi.$$

Next, we make the change of variables $y = u - \ln b(V,\tau)$ and obtain

$$W(x,V,\tau) = \int_0^\tau \int_0^\infty \int_{-\infty}^\infty H(y)e^{r\xi}(qe^{y+\ln b(w,\xi)} - rK) \times$$
$$\times \left\{\frac{1}{2\pi}\int_{-\infty}^\infty e^{-i\phi(y+\ln b(w,\xi))}\hat{f}_2(x,v,\tau-\xi,\phi,w)d\phi\right\} dy\, dw\, d\xi.$$

Now substitute $\ln b(V,\tau)$ from equation (1.8) and obtain an expression for $W(x,V,\tau)$:

$$W(x,V,\tau) = \int_0^\tau \int_0^\infty \int_{-\infty}^\infty H(y) e^{r\xi} (q e^{y+b_0(\tau)+Vb_1(\tau)} - rK) \times$$

$$\times \left\{ \frac{1}{2\pi} \int_{-\infty}^\infty e^{-i\phi(y+b_0(\tau)+Vb_1(\tau))} \hat{f}_2(x,v,\tau-\xi,\phi,w) d\phi \right\} dy\, dw\, d\xi. \quad (1.9)$$

To bring this expression for $W(x,V,\tau)$ to the semi-analytic form we used a proof from [9]. Then we obtain the final expression:

$$W(x,V,\tau) = e^x e^{r\tau} \int_0^\tau q e^{-q(\tau-\xi)} \left( \frac{1}{2} + \frac{1}{\pi} \int_0^\infty \mathrm{Re}\left( \frac{e^{-i\phi b_0(\tau)}}{i\phi} f_1(x,v,\tau-\xi,\phi,-b_1(\xi)\phi) \right) d\phi \right) d\xi -$$

$$- \int_0^\tau rK e^{-r(\tau-\xi)} \left( \frac{1}{2} + \frac{1}{\pi} \int_0^\infty \mathrm{Re}\left( \frac{e^{-i\phi b_0(\tau)}}{i\phi} f_2(x,v,\tau-\xi,\phi,-b_1(\xi)\phi) \right) d\phi \right) d\xi$$

$$f_1(x,v,\tau-\xi,\phi,\psi) \equiv e^{-x} e^{-(r-q)(\tau-\xi)} f_2(x,v,\tau-\xi,\phi-i,\psi)$$

Returning to equation (1.6), all that is left to determine is the value $U(x,V,\tau)$ of a European option with stochastic volatility. As mentioned above, the solution of this problem can be found in Heston (1993). Here is an expression without proof:

$$U(x,v,\tau) = e^x e^{(r-q)\tau} \left( \frac{1}{2} + \frac{1}{\pi} \int_0^\infty \mathrm{Re}\left( \frac{e^{-i\phi \ln K}}{i\phi} f_1(x,v,\tau,\phi,0) \right) d\phi \right) -$$

$$- K \left( \frac{1}{2} + \frac{1}{\pi} \int_0^\infty \mathrm{Re}\left( \frac{e^{-i\phi \ln K}}{i\phi} f_2(x,v,\tau,\phi,0) \right) d\phi \right)$$

Let us combine the results for the value $U(x,V,\tau)$ of a European option and the early exercise premium $W(x,V,\tau)$ of an American option, as well as take into account the change of variables made above $S = e^x$, $C_A(S,V,\tau) = e^{-r\tau} U(x,V,\tau)$. As a result, we obtain the final expression for the price of American Call option with vanilla payoff function:

$$C_A(S,V,\tau) = Se^{-q\tau} P_1(S,V,\tau,K,0) - Ke^{-r\tau} P_2(S,V,\tau,K,0) +$$

$$+ \int_0^\tau qSe^{-q(\tau-\xi)} P_1(S,V,\tau-\xi,e^{b_0(\xi)},-b_1(\xi)) d\xi - \quad (1.10)$$

$$- \int_0^\tau rKe^{-r(\tau-\xi)} P_2(S,V,\tau-\xi,e^{b_0(\xi)},-b_1(\xi)) d\xi$$

where

$$P_j(S,V,\tau,K,\psi) = \frac{1}{2} + \frac{1}{\pi} \int_0^\infty \mathrm{Re}\left( \frac{e^{-i\phi \ln K}}{i\phi} f_j(\ln S,V,\tau,\phi,\psi) \right) d\phi \quad (1.11)$$

All that is left now is to get the function surface early performance and present it in the form of a linear approximation $\ln b(V, \tau) \approx b_0(\tau) + V b_1(\tau)$. The idea of constructing this surface is described in the next section.

## Costruction of the early exercise surface

The idea of constructing the early exercise surface was proposed in the article Longstaff & Schwartz (2001) [11]. To implement this idea, we introduce a probability space $(\Omega, F, P)$ and finite time horizon [0, T]. Here the space of elementary events $\Omega$ is a set of all possible realizations of a stochastic process $(S_t, V_t)$ on the interval [0, T]. The elementary events we denote as $\omega$, which is a sample of a trajectory. The quantities $F, P$ are, respectively, the probability filtering and probability measure. Assume the existence of equivalent martingale measure for this stochastic process. A consequence of the existence of a martingale measure is the lack of arbitrage opportunities in this economy. We restrict our attention to options, for which payoffs are elements of the space $L^2(\Omega, F, P)$. This gives us reason to decompose the option payoff in the functional basis in the space $L^2(\Omega, F, P)$. In our case we choose a basis of the Laguerre polynomials.

The obvious way to assess the price of American option is to take it equal to the maximal expected discounted cash flow, where the maximum is taken over all possible exercise times of option. The expectation is calculated by the martingale (risk-neutral) measure. Thus, we are interested in obtaining the optimal option exercise strategy that maximizes the expected cash flow. Since the option cash flow at any time depends on the price of the underlying asset, which is a random variable, then cash flow is also a random variable. Assume that the option can be exersied only in a finite set of time $0 = t_0 < t_1 < t_2 < \ldots < t_N = T$ located in the interval [0, T]. Then it is necessary to generate a set of $M$ different trajectories of basic stochastic processes $(S_t, V_t)$ containing $N$ time steps. At time $T$ the expected cash flow is simply equal to the payoff function. Consequently, the optimal strategy for the option holder is to exercise option, if the payoffs function greater than zero. If the option holder decides not to exercise the option at time $t_k < T$ than the expected discounted (to time $t_k$) cash flow will be equal

$$H(\omega_j, t_k) = E\left(\sum_{i=t_{k+1}}^{T} e^{r(t_i - t_k)} C(\omega_j, t_i)\right),$$

where $C(\omega_j, t_i)$ is a cash flow, which will occur at time $t_i$.

To calculate $H(\omega_j, t_k)$ for all trajectories, we approximate the function $H(\omega_j, t_k)$ with the help of polynomials. The choice of these polynomials can be debated, but here we use the Laguerre polynomials. The expansion coefficients will be determined by the method of least squares:

$$H(S_k, V_k, t_k) \approx C_0^k + C_1^k L_0\left(\frac{S}{K}\right) + C_2^k L_1\left(\frac{S}{K}\right) + C_3^k L_0\left(\frac{S}{K}\frac{V}{Q}\right),$$

were $L_0, L_1$ are the Laguerre polynomials of degree 0 and 1 respectively. Polynomial expansion is possible due to the fact that we have assumed that the payoff functions belong to $L^2(\Omega, F, P)$. Now, if the option holder follows the optimal strategy, it is necessary to compare the instantaneous cash flow $C(\omega_j, t_k)$, which will happen if the option to exercise at the time $t_k$ and the expected future discounted cash flow $H(\omega_j, t_k)$. Thus, if $C(\omega_j, t_k) > H(\omega_j, t_k)$, the option has to exersice, otherwise, one need to decide to keep the option until the next discrete point in time.

The decision to exercise or not exercise the option needs to be done for each time $t_k$. As a result, we obtain the unique optimal time when the option is exercised. If one executes an option at this time, he will receive maximum discounted cash flow. All that remains is to average over many different trajectories of the stochastic process describing the dynamics of the underlying asset. So the easiest way to evaluate American options is generating trajectories by means of the Monte Carlo method. Here we want to obtain an analytical expression for the price of American option and all that we need to do is to evaluate the early exercise surface. To do this, it is necessary for every time $t_k$ and for each possible value of $V_k$ determine the value of the underlying asset $S_k$, which satisfied the boundary condition for the Call option:

$$C_A(b(V,\tau), V, \tau) = b(V,\tau) - K.$$

Once we get the price of the underlying asset, in which for different values of volatility exercise of the option is optimal, we can define functions $b_0(\tau), b_1(\tau)$ using the method of least squares, on the basis of the expression

$$\ln b(V,\tau) \approx b_0(\tau) + V b_1(\tau).$$

Now we have prepared everything necessary in order to calculate the price of an American call option, using equation (1.10).

## Numerical calculations and results

To generate the trajectories of the underlying asset and its volatility we shall use the Euler scheme:

$$S_{t+1} = S_t + \mu S_t \Delta t + \sqrt{V_t} S_t \Delta W_1$$
$$V_{t+1} = V_t + k(Q - V_t)\Delta t + \sigma \sqrt{V_t} \left( \rho \Delta W_1 + \sqrt{1-\rho^2} \Delta W_2 \right)$$

where $\Delta W_1, \Delta W_2$ are independent random variables distributed with the parameters $(0, \sqrt{\Delta t})$

To calculate the integrals in equation (1.10) and equation (1.11), we used the quadrature formulas, namely, the trapezoidal formula.

Table 1 compares the results obtained by the method described in this paper and the the results obtained by the method derived in the article AitSahlia and Lai (1999). The column $T$ represents time until the option is exercised, the column $S$ represents the price of underlying asset, the column "Test" represents the results for the American option prices with which we compared our results. Columns Mkn represent our results. The abbreviation MkN denotes prices obtained using $M \times 1,000$ sample paths and $N$ time steps.

| T | S | Test | 10k*10 | 50k10 | 100k5 | 100k10 | 100k25 |
|---|---|---|---|---|---|---|---|
| 0.02 | 80 | 20.002 | 19.98 | 19.98 | 19.96 | 19.98 | 19.992 |
|  | 90 | 10 | 9.937 | 9.958 | 9.952 | 9.941 | 9.927 |
|  | 100 | 2.138 | 2.114 | 2.112 | 2.121 | 2.112 | 2.106 |
|  | 120 | 0.001 | 0.001 | 0.001 | 0.001 | 0.001 | 0.001 |
| 0.08 | 80 | 20 | 19.943 | 19.961 | 19.895 | 19.947 | 19.969 |
|  | 90 | 10.354 | 10.408 | 10.487 | 10.501 | 10.428 | 10.356 |
|  | 100 | 3.942 | 3.971 | 3.974 | 4.007 | 3.964 | 3.937 |
|  | 120 | 0.19 | 0.223 | 0.223 | 0.227 | 0.223 | 0.221 |
| 0.15 | 80 | 20.002 | 20.007 | 20.052 | 19.98 | 20.017 | 20.006 |
|  | 90 | 10.842 | 11.007 | 11.072 | 11.15 | 11.012 | 10.893 |
|  | 100 | 4.979 | 5.049 | 5.043 | 5.116 | 5.038 | 4.979 |
|  | 120 | 0.651 | 0.66 | 0.659 | 0.671 | 0.659 | 0.65 |

Table1. Comparison of results obtained in present paper and the results of the article AitSahlia and Lai (1999)

Table 1 show that the method of modeling the early exercise surface by the Monte Carlo method gives fair results, and may be useful in practice. To implement the method described in this paper it has been written the software in the language of Matlab package.

## Conclusion

In this paper, a method for evaluating American options under stochastic volatility models [9,10,11] has been developed. In particular, there was developed an algorithm that simulates the early excersise surface of American options and used the idea [9] of obtaining analytical expressions for the value of options of American type with vanilla payoff function. The results showed that the method described in the paper allows to accurately calculating the price of American options. As a logical continuation of this study it may be interesting to develop a method for assessing American options for the complex payoff functions, using the jump diffusion process for describing the dynamics of the underlying asset.